\documentclass[conference]{IEEEtran}
\usepackage{cite}
\usepackage{amsmath,amssymb,amsfonts}
\usepackage{algorithmic}
\usepackage{graphicx}
\usepackage{textcomp}
\usepackage{hyperref}
\usepackage{xcolor}
\usepackage{moresize}
\usepackage{placeins}
\usepackage{orcidlink}
\hypersetup{
    colorlinks,
    linkcolor={blue!80!black},
    citecolor={blue!80!black},
    urlcolor={blue!80!black}
}
\def\BibTeX{{\rm B\kern-.05em{\sc i\kern-.025em b}\kern-.08em
    T\kern-.1667em\lower.7ex\hbox{E}\kern-.125emX}}
\begin{document}

\title{Entangled happily ever after: Wedding reception seating mapped to classical and quantum optimizers\\
% \title{Until wavefunction collapse do we part: Wedding seating optimization as an \texttt{NP}-hard problem solvable on quantum computers\\
% \thanks{KAN is funded by the Foundation for Health Care Quality.  VKM is funded by the Simons Foundation.}
}

\author{\IEEEauthorblockN{Karie A. Nicholas \orcidlink{0009-0005-3233-6689}}
\IEEEauthorblockA{\textit{Foundation for Health Care Quality}\\
Seattle, WA, United States of America \\
\href{mailto:knicholas@qualityhealth.org}{knicholas@qualityhealth.org}}
\and
\IEEEauthorblockN{Vikram Khipple Mulligan \orcidlink{0000-0001-6038-8922}}
\IEEEauthorblockA{\textit{Center for Computational Biology, Flatiron Institute}\\
New York, NY, United States of America \\
\href{mailto:vmulligan@flatironinstitute.org}{vmulligan@flatironinstitute.org}}
}

\maketitle

\begin{abstract}
    Although optimization is one of the most promising applications of quantum computers, the development of effective optimization strategies requires real-world test cases.  When planning our recent wedding reception, we realized that the problem of optimally seating our guests, given constraints related to guests' relatedness, shared interests, and physical needs, could be mapped to a cost function network (CFN) form solvable with classical or quantum optimization algorithms.  We compared the seating optimization performance of classical Monte Carlo CFN solvers in the Masala software suite to that of quantum annealing-based CFN optimization algorithms using one-hot, domain-wall, and approximate binary mappings, which we had developed for protein design problems.  Surprisingly, the D-Wave Advantage 2 system, which performs well on similarly-structured CFN problems for protein design, struggled to return optimal seating arrangements that were easily found by classical Monte Carlo methods.  We provide our seating optimization benchmark set, and code to convert seating optimization problems to CFN problems, as a plugin library for Masala, permitting this class of real-world problems to be used to benchmark performance of current and future classical CFN solvers, quantum optimization algorithms, and quantum computing hardware.
\end{abstract}

\begin{IEEEkeywords}
quantum computing, quantum annealing, optimization, benchmarking, cost function network problems, Monte Carlo methods, Masala software suite
\end{IEEEkeywords}

\section{Introduction}

    On 31 October 2025, the authors were married. We were faced with the difficult task of optimally seating our guests for the supper reception in order to satisfy many constraints pertaining to degrees of relatedness, compatibility of personalities, and guests' physical needs in order to maximize guests' enjoyment and minimize discord. As one does when planning one's wedding, we sought to map the optimization problem to a known form amenable to solution either with classical or quantum optimization algorithms.
    
    The seating optimization problem may be formulated as follows: given $D$ seats and $N$ guests (with $D \geq N$), find a vector $\dot{\vec{s}}=\left(\dot{s}_0,\dot{s}_1,,\dots,\dot{s}_{N-1}\right)$ where each $\dot{s}_i \in \left[0,D-1\right]$ assigns one seat to each guest, ensuring that no two guests are assigned the same seat, subject to the requirement that an objective function $f\left(\vec{s}\right)$ is minimized.  (We use a dot here to indicate an optimal assignment.)  The objective function $f\left(\vec{s}\right)$ encodes the various constraints that must be satisfied for optimal seating.  As stated, this problem has $D^N$ possible solutions.  The constraint that no two guests may be assigned the same seat reduces the set of allowed solutions to $\frac{D!}{\left(D-N\right)!}$ possibilities (which is less than $D^N$).  Penalties or bonuses may be assigned either based on a single guest's seating preferences (\textit{e.g.} a penalty may be assigned if a guest bothered by noise is seated near a speaker) or based on the interactions between pairs of guests (\textit{e.g.} bonuses may be assigned for couples sitting together, penalties may be assigned to keep pairs with conflicting personalities apart, \textit{etc.}).  Among the pairwise penalties, constraints related to mitigating the spread of infectious disease to vulnerable individuals may also be included -- a consideration at our own wedding.
    
    The seating optimization problem is related to the problem of optimizing amino acid sequences when designing proteins or other heteropolymers (reviewed in \cite{ghodge_computational_2022}). Both problem types can be mapped to a class of \texttt{NP}-hard mathematical problems called cost function network optimization (CFN) problems \cite{pierce_protein_2002}.  In past work, we showed that CFN problems, typically solved on classical hardware by dead-end elimination \cite{leach_exploring_1998}, branch-and-bound \cite{traore_new_2013}, or Monte Carlo methods \cite{kuhlman_native_2000,zaborniak_open-source_2025-1}, may also be solved on current-generation quantum annealers using a one-hot encoding \cite{mulligan_designing_2019}.  More recently, we have tested a domain-wall encoding described by Chancellor \cite{chancellor_domain_2019}, and an approximate binary encoding that we and our collaborators developed (manuscript in preparation and \cite{zaborniak_toward_2025}).  We also recently introduced the Masala software suite, a library that extends existing software (such as the Rosetta software suite, commonly used for protein design \cite{leaver-fay_rosetta3_2011,leman_macromolecular_2020}) with plugin optimizers, to facilitate the use of new optimization methods in the context of existing pipelines.  Here, we show that the difficult problem of optimizing seating at a wedding reception or other gathering can be mapped to a CFN problem, and can consequently be solved using classical or quantum optimizers.

    The present work provides three useful contributions to the optimization field.  First, it provides a Masala plugin library and an input format for automatically converting seating optimization problems to CFN problems, and for solving them using any solver supported by the Masala software suite's plugin libraries at present or in the future.  Second, it introduces a set of benchmark seating optimization problems that can be used to evaluate the performance of classical or quantum optimization algorithms.  We demonstrate these problems as a benchmark for a classical Monte Carlo algorithm, for an experimental hill-flattening variant, and for quantum annealing-based optimization on the D-Wave Advantage 2 system using three different mappings (one-hot, domain-wall, and approximate binary), and discuss interesting revelations about the relative performance of classical and quantum solvers.  And third, it provides means of solving a real-world problem understandable to laypeople on current and future classical or quantum hardware, one with practical applications from event planning to resource allocation in infection prevention settings.

\section{Methods}

\subsection{The Seating Optimization Masala Plugins library}

\begin{figure}
    \centering
    \includegraphics[]{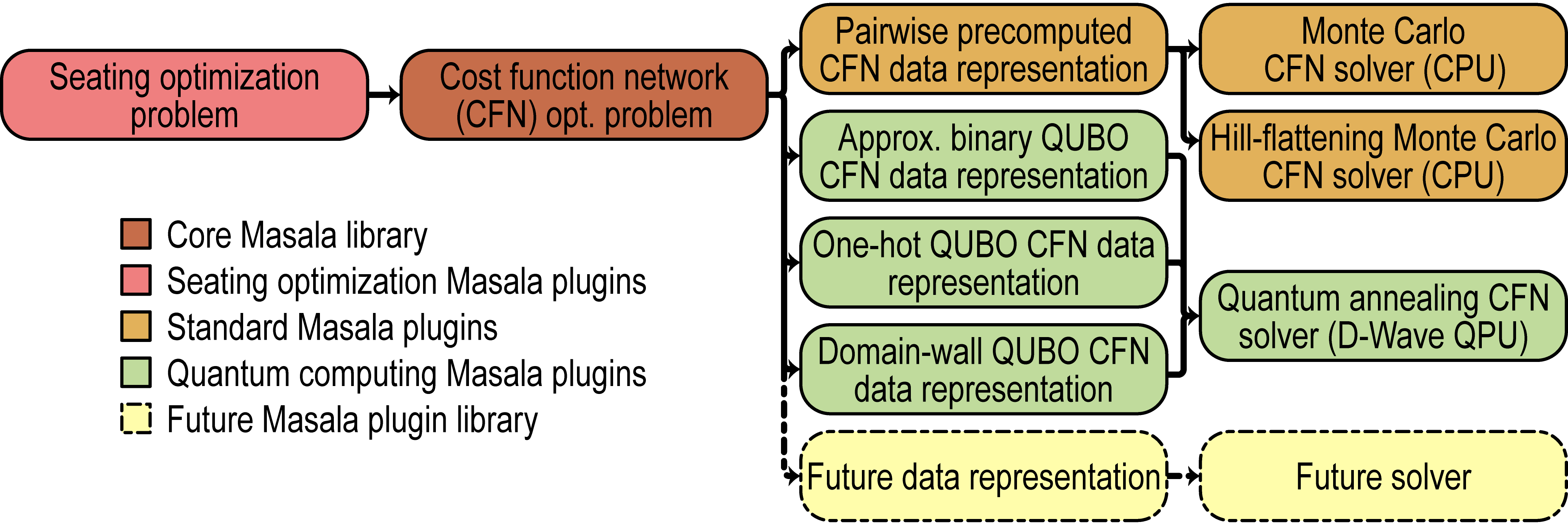}
    \caption{Function of the \begin{scriptsize}\texttt{optimize\_seating}\end{scriptsize} application in the Seating Optimization Masala Plugins library.  This application permits a user to configure a seating optimization problem (salmon), converts it into a general CFN problem (brown), and encodes it as any compatible data representation available in a Masala plugin library (orange, green), permitting the optimization to be performed with any plugin Masala solver.  New data representations and solvers employing new algorithms or new hardware may be loaded as they are developed (yellow).}
    \label{fig:fig_diagram}
\end{figure}

    The Masala software suite is a set of C\texttt{++} libraries that we originally developed for heteropolymer design, validation, and analysis \cite{zaborniak_open-source_2025-1}.  Its modular, plugin-based architecture permits easy recombination of its optimization tools to apply them to new classes of problems.  The Standard Masala Plugins library includes an efficient, parallelized Monte Carlo CFN problem solver.  We are developing a hill-flattening variant, which adaptively alters the objective function being optimized based on the lowest-scoring solution yet encountered to flatten barriers between local optima. A Quantum Computing Masala Plugins library has also been developed and is also planned for release, providing one-hot, domain-wall, and approximate binary encodings of CFN problems for solution on quantum annealers (manuscript in preparation).  For this work, we created the Seating Optimization Masala Plugins library, providing easy means of converting banquet layout descriptions and guest needs constraints into a CFN problem, and then solving it using any current or future Masala plugin CFN optimizer (\textbf{Fig. \ref{fig:fig_diagram}}).

    Masala's core and Standard Plugins libraries are available from \href{https://github.com/flatironinstitute/masala_public}{https://github.com/flatironinstitute/masala\_public} and \href{https://github.com/flatironinstitute/masala_public_standard_plugins}{https://github.com/flatironinstitute/masala\_public\_standard\_- plugins}, respectively.  The new Seating Optimization Masala Plugins library is available from \href{https://github.com/flatironinstitute/masala_public_seating_optimization_plugins}{https://github.com/flatironinstitute/masala\_public\_seating\_- optimization\_plugins}.  All of these are made available under a free-and-open-source AGPL 3.0 licence.

\subsection{Encoding seating constraints as a CFN problem}

    A CFN problem is defined by a set of $D_i$ choices for each of $N$ nodes (where $i \in [0,N-1]$), and a cost function $f(\vec{s})$, where $\vec{s}$ is a candidate solution vector of length $N$ in which each entry represents the index of the discrete choice chosen for each node.  The objective is to find the optimal solution $\dot{\vec{s}}$ that minimizes $f(\vec{s})$.  In the special case of a \textit{pairwise-decomposable} CFN problem, $f(\vec{s})$ has the form given by Eq. \eqref{eq:cfn_pairwise}.

    \begin{equation}
    f\left(\vec{s}\right) = \sum\limits_{i=0}^{N-1} \alpha^{(i)}\left(s_i\right) + \sum\limits_{i=0}^{N-2}\sum\limits_{j=i+1}^{N-1} \beta^{(i,j)}\left(s_i,s_j\right)
    \label{eq:cfn_pairwise}
    \end{equation}

    In Eq. \eqref{eq:cfn_pairwise}, $\alpha^{(i)}(s_i)$ and $\beta^{(i,j)}(s_i,s_j)$ are functions of the choice at a particular node $i$, and of the pair of choices at a particular pair of nodes $(i,j)$, respectively.  These functions' values are generally exhaustively enumerable and amenable to classical precomputation.
    
    When seating guests, one may either interpret seats as choices and guests as nodes, or the converse.  Since it is possible to have more seats than guests, we chose the former.  We developed an input file format allowing a user to define the number and layout of seats, and allowing Cartesian coordinates and orientation of seats to be computed.  We also allowed the user to restrain subsets of guests to subsets of the seats, to simplify the optimization problem.  We then encoded four types of user-defined seating constraints as two-node penalties $\beta^{(i,j)}(s_i,s_j)$.  In our code, we defined a \texttt{Constraint} pure virtual base class from which derived classes for particular seating constraints inherited, permitting additional constraint types to be added in the future.

\subsubsection{No two guests may be assigned the same seat}

    The \texttt{GuestOverlapConstraint} class implements this constraint by adding to the two-node penalties a positive penalty for the same choice at two nodes (\textit{i.e.} the same seat to be assigned to two guests):

    \begin{equation}
    \beta^{(i,j)}_{\text{overlap}}\left( s_i, s_j \right) = \delta\left(s_i,s_j\right) p_\text{overlap}
    \label{eq:overlap_constraint}
    \end{equation}

    In Eq. \eqref{eq:overlap_constraint} and subsequent equations, $\delta\left(s_i,s_j\right)$ is the Kronecker delta function (which is $1$ if $s_i=s_j$ and 0 if $s_i \ne s_j$) and $N$ is the number of guests. The penalty $p_\text{overlap}$ is positive.

\subsubsection{It may be favourable or unfavourable for a particular pair of guests to be in adjacent seats, or at the same table}
\label{subsubsec:adjacent_pair_constraint}

    The \texttt{GuestPairAdjacentSeatConstraint} and \texttt{GuestPairSameTableConstraint} classes constrain the seating of guest pairs at adjacent seats and at the same table, respectively.  These take $N_\text{gpair}$ user-defined penalties $\left\{p^{(0)}_{\text{gpair}},\dots,p^{(N_\text{gpair}-1)}_{\text{gpair}}\right\}$ for $N_\text{gpair}$ user-defined guest pairs $\left\{\left(g^{(0)}_{0},g^{(0)}_{1}\right),\dots, \left(g^{\left(N_\text{gpair}-1\right)}_{0},g^{\left(N_\text{gpair}-1\right)}_{1}\right)\right\}$.  They also take a vector of $N_\text{spair}$ seat pairs, $\vec{\sigma} = \left\{\left(\sigma^{(0)}_{0},\sigma^{(0)}_{1}\right), \dots, \left(\sigma^{\left(N_\text{spair}-1\right)}_{0},\sigma^{\left(N_\text{spair}-1\right)}_{1}\right)\right\}$, that are either adjacent or at the same table.  The penalty function added to the two-node penalties is given by Eq. \eqref{eq:guest_pair_constraint}:

    \begin{equation}
    \scalebox{0.78}{$
    \beta^{\left(i,j\right)}_\text{adj/table}\left(s_i,s_j\right) =
    \begin{cases}
        \sum\limits^{N_\text{gpair}-1}_{k=0} \delta\left(i,g^{\left(k\right)}_{0}\right)  \delta\left(j,g^{\left(k\right)}_{1}\right) p^{(k)}_\text{gpair} & \text{if}~\left(s_i,s_j\right) \in \vec{\sigma} \\
        0 & \text{otherwise}
    \end{cases}
    $}
    \label{eq:guest_pair_constraint}
    \end{equation}

\subsubsection{It may be favourable or unfavourable for a pair of guests to be seated in close spatial proximity}

    Since Cartesian coordinates of seats are fixed and are evaluated classically prior to setting up the CFN problem, constraints based on geometry, such as classically-computed distances between seats, are possible.  For our purposes, we used Gaussian functions given by Eq. \eqref{eq:gaussian} to encourage or discourage spatial proximity between guests.
    
    \begin{equation}
    G^{(i)}\left(s_a,s_b\right) = \frac{p^{(i)}_\text{prox}}{e^{-1/\left(\lambda^{(i)}\right)^2}}e^{-\left(\frac{d\left(s_a,s_b\right)}{\lambda^{(i)}}\right)^2}
    \label{eq:gaussian}
    \end{equation}

    In Eq. \eqref{eq:gaussian}, $i$ is the index of the guest pair to constrain, $p^{(i)}_\text{prox}$ represents the penalty value at a separation of one unit, $\lambda^{(i)}$ is the Gaussian breadth, and $d\left(s_a,s_b\right)$ is the classically-precomputed distance between seats $s_a$ and $s_b$ given by Eq. \eqref{eq:cartdist}:

    \begin{equation}
    d\left(s_a,s_b\right) = \sqrt{\left(x_a-x_b\right)^2+\left(y_a-y_b\right)^2}
    \label{eq:cartdist}
    \end{equation}

% $N_\text{gpair}$ user-defined penalties $\left\{p^{(0)}_{\text{gpair}},\dots,p^{(N_\text{gpair}-1)}_{\text{gpair}}\right\}$ for $N_\text{gpair}$ user-defined guest pairs $\left\{\left(g^{(0)}_{0},g^{(0)}_{1}\right),\dots, \left(g^{\left(N_\text{gpair}-1\right)}_{0},g^{\left(N_\text{gpair}-1\right)}_{1}\right)\right\}$.

    Given a set of $N_\text{gpair}$ guest pairs $\left\{\left(g^{(0)}_{0},g^{(0)}_{1}\right),\dots, \left(g^{\left(N_\text{gpair}-1\right)}_{0},g^{\left(N_\text{gpair}-1\right)}_{1}\right)\right\}$ to constrain with a set of $N_\text{gpair}$ Gaussian proximity functions $\left\{ G^{(0)} \left(s_a,s_b\right), \dots, G^{\left(N_\text{gpair}-1\right)} \left(s_a,s_b\right) \right\}$, the penalty that is added to the two-node penalties is given by Eq. \eqref{eq:prox_constraint}:

    \begin{equation}
    \scalebox{.98}{$
    \beta^{\left(i,j\right)}_\text{prox}\left(s_i,s_j\right) =
        \sum\limits^{N_\text{gpair}-1}_{k=0} \delta\left(i,g^{\left(k\right)}_{0}\right)  \delta\left(j,g^{\left(k\right)}_{1}\right)G^{(i)}\left(s_j,s_k\right)
    $}
    \label{eq:prox_constraint}
    \end{equation}

    Given Eqs. \eqref{eq:overlap_constraint}, \eqref{eq:guest_pair_constraint}, and \eqref{eq:prox_constraint}, and setting $\alpha^{(i)}\left(s_i\right)=0~\forall~i$ (in the absence of any current constraints dependent on a single guest's seat assignment), we can rewrite Eq. \eqref{eq:cfn_pairwise} as:

    \begin{equation}
    %\scalebox{.69}{$
    \begin{split}
        f\left(\vec{s}\right) = \sum\limits_{i=0}^{N-2}\sum\limits_{j=i+1}^{N-1} & \beta^{(i,j)}_\text{overlap}\left(s_i,s_j\right) + \beta^{(i,j)}_{\text{adj}}\left(s_i,s_j\right) \\ & + \beta^{(i,j)}_{\text{table}}\left(s_i,s_j\right) + \beta^{(i,j)}_{\text{prox}}\left(s_i,s_j\right)
    \end{split}
    %$}
    \label{eq:final_penalty_fxn}
    \end{equation}

\subsection{Test problems}

    The \texttt{test\_data/} sub-directory of the Seating Optimization Masala Plugins repository contains five problems ranging from easy to hard, which may be used to test the capabilities of classical or quantum solvers.  The input file format, documented in \texttt{README.md}, makes it easy to define additional seating optimization problems.

\subsubsection{``Four friends and enemies''}

    This extremely simple test case (\texttt{test\_data/example\_problem\_1.txt}) involves four people seated around a round table.  Three pairings (Avery and Basil, Avery and Charlie, and Charlie and Dawson) are amicable, but a fourth (Basil and Charlie) is marked by enmity.  Friends may be seated next to one another, but enemies should not.  By inspection, the optimal solutions are the eight permutations that place Basil and Charlie opposite one another, with Avery and Dawson in either arrangement in the remaining two seats.

\subsubsection{``The odd man out''}

    This fairly simple problem (\texttt{test\_data/example\_problem\_2.txt}) involves five people to be divided over a four-person table and a one-person table.  Four of the people have pairwise friendships.  The fifth is friends with two, but enemies with the other two.  It is favourable to seat friends next to one another, unfavourable to seat enemies next to one another, and somewhat unfavourable to seat enemies at the same table.  This yields eight optimal configurations, all with the fifth person at the solo table and the other four arranged clockwise or counterclockwise around the four-person table.

\subsubsection{``The unctuous uncle''}

    In this example (\texttt{test\_data/example\_problem\_3.txt}), we imagined a seventeen-person wedding banquet with many of the usual constraints (couples should sit together, the parents of the married couple should be at their table, siblings should preferably be at the same table), but added a twist: one uncle is an unpleasant womanizer who ought to be kept far away from those women whom he might harass.  We sought to discourage seating arrangements that put the imaginary uncle adjacent to, at the same table as, or in spatial proximity to any potential victim of his unctuousness.

\subsubsection{``The Capulets \& Montagues''}

    This scenario (\texttt{test\_data/example\_problem\_4.txt}) imagines the wedding feast of Romeo and Juliet had both families (25 people) been invited.  There are five tables, four of which have five seats each, while the head table (to which the bride and groom and their parents are restrained) has six.  One seat at one of the other four tables will inevitably be empty.  Constraints encourage couples (\textit{e.g.} Lady and Lord Capulet, Lady and Lord Montague) to be seated together and friends (\textit{e.g.} Gregory and Sampson, Benvolio and Mercutio) to be adjacent, at the same table, or in spatial proximity.  Penalties discourage characters with the greatest enmity (\textit{e.g.} hotheads Tybalt and Mercutio, Romeo and former flame Rosaline) from sitting next to one another, at the same table, or in spatial proximity.  While the optimal solution cannot be simply discovered, it is expected that good solutions will largely keep the houses of Capulet and Montague separate, with particular focus on separating the most pugilistic characters.

\subsubsection{The Nicholas-Mulligan wedding}
    The seating problem for our own 37-person wedding reception, with guests anonymised, is defined in \texttt{test\_data/example\_problem\_5.txt}.  As before, we encoded bonuses ensuring that couples and families sit together, and assigned penalties or bonuses for guest pairings that we thought would conflict or be favourable based on our assessment of our guests' personalities -- for instance, encouraging the seating of two guests who were scientists at the same table as a child enthusiastic about science.  Since two of our guests were native French speakers, we also assigned bonuses to having an English speaker who was proficient in French at the same table and in close proximity.  One set of disease transmission constraints was also introduced: since we had one guest who could not receive a COVID-19 booster due to medical reasons, we constrained this guest to be seated as far as possible from all guests susceptible to adverse outcomes (\textit{i.e.} those elderly or otherwise medically compromised).  We simplified the problem slightly by assigning ourselves to two seats, and six close family members of our choosing to our table, but permitted the seating order of the six guests at this table, and the assignment of all other guests to the other tables, to be determined by the solver.  We also devised a further-simplified version, \texttt{test\_data/example\_problem\_5\_simplified.txt}, that added additional restraints for a second group of guests restricted to a central table, allowing this problem to be mapped to the number of qubits available on current QPUs.

\section{Results}

    For each of the six test problems (including the simplified variant of problem 5, ``Pr. 5s''), the scores of the top solutions found by each of several solvers is shown in \textbf{Table \ref{tab:best_scores}}, with the number of guest overlaps (guests assigned to the same seat) shown in round brackets.  Lower is better for both, and the lowest found by any method is shown in bold for each problem.  Score value comparisons are only meaningful within a problem.  We tested the Standard Masala Plugins library's classical, CPU-based \texttt{MonteCarloCostFunctionNetworkOptimizer} (``MC'' in \textbf{Table \ref{tab:best_scores}}) \cite{zaborniak_open-source_2025-1} and experimental \texttt{HillFlatteningMonteCarloCostFunctionNet- workOptimizer} (``HF''), allowing trajectories of 1,000 steps (``1k'') 30,000 steps, (``30k''), 1,000,000 steps (``1M''), or 1,000,000 steps performed in parallel on each of 32 threads (``32M'').
    We also tested the Quantum Computing Masala Plugins library's \texttt{DWaveQuantumQUBOProblemOptimizer} (manuscript in preparation), running on the D-Wave Advantage 2 system 1 quantum annealer.  (Prior to our wedding, we attempted to run problems on the Advantage 2 system 1.6, available at the time. Later, we performed additional benchmarks on system 1.13).  We used three data representations: a one-hot QUBO encoding (``OH'') that uses $ND$ logical qubits for problems with $N$ guests and $D$ seats \cite{mulligan_designing_2019}, a domain-wall QUBO encoding (``DW'') \cite{chancellor_domain_2019} that uses $N\left(D-1\right)$ logical qubits, and an approximate binary QUBO encoding (``AB'') that uses $N\lceil \text{log}_2 D\rceil$ logical qubits (manuscript in preparation and \cite{zaborniak_toward_2025}).  We embedded problems using the \texttt{minor\_miner} algorithm \cite{cai_practical_2014}, and in the case of the approximate binary encoding, we used inhomogeneous driving to improve solution quality \cite{adame_inhomogeneous_2020}.  Sampling runs of 1,000 (``1k''), 30,000 (``30 k''), or 1,000,000 (``1M'') shots were carried out.

\begin{table}[]
    \centering
    \begin{tabular}{p{8mm}| c c c c c c}
        \centering{\scriptsize{Solver}} & \scriptsize{Prob. 1} & \scriptsize{Prob. 2} & \scriptsize{Prob. 3} & \scriptsize{Prob. 4} & \scriptsize{Prob. 5s} & \scriptsize{Prob. 5} \\
        \hline
        \hline
        %OHQA 30k & ** & ** & ** & ** & ** & ** \\
        \centering{\ssmall{OH 1M}} & \ssmall{NVS} & \ssmall{---} & \ssmall{---} & \ssmall{---} & \ssmall{---} & \ssmall{---} \\
        %DWQA 30k & ** & ** & ** & ** & ** & ** \\
        \centering{\ssmall{DW 1M}} & \ssmall{NVS} & \ssmall{---} & \ssmall{---} & \ssmall{---} & \ssmall{---} & \ssmall{---} \\
        \centering{\ssmall{AB 1k}} & \ssmall{\textbf{-15.0(0)}} & \ssmall{-5.0(2)} & \ssmall{136.1(7)} & \ssmall{250.8(14)} & \ssmall{106.3(17)} & \ssmall{NE} \\
        \centering{\ssmall{AB 30k}} & \ssmall{\textbf{-15.0(0)}} & \ssmall{-5.0(2)} & \ssmall{121.4(8)} & \ssmall{326.2(13)} & \ssmall{160.2(22)} & \ssmall{NE} \\
        \centering{\ssmall{AB 1M}} & \ssmall{\textbf{-15.0(0)}} & \ssmall{-25.0(2)} & \ssmall{116.5(8)} & \ssmall{456.8(14)} & \ssmall{198.9(22)} & \ssmall{NE} \\
        \hline
        \centering{\ssmall{MC 1k}} & \ssmall{\textbf{-15.0(0)}} & \ssmall{\textbf{-40.0(0)}} & \ssmall{\textbf{-64.9(0)}} & \ssmall{-104.1(0)} & \ssmall{-465.9(0)} & \ssmall{-499.1(0)} \\
        \centering{\ssmall{MC 30k}} & \ssmall{\textbf{-15.0(0)}} & \ssmall{\textbf{-40.0(0)}} & \ssmall{\textbf{-64.9(0)}} & \ssmall{-114.6(0)} & \ssmall{-479.5(0)} & \ssmall{-501.0(0)} \\
        \centering{\ssmall{MC 1M}} & \ssmall{\textbf{-15.0(0)}} & \ssmall{\textbf{-40.0(0)}} & \ssmall{\textbf{-64.9(0)}} & \ssmall{-113.8} & \ssmall{-505.2(0)} & \ssmall{-511.1(0)} \\
        \centering{\ssmall{MC 32M}} & \ssmall{\textbf{-15.0(0)}} & \ssmall{\textbf{-40.0(0)}} & \ssmall{\textbf{-64.9(0)}} & \ssmall{\textbf{-123.7(0)}} & \ssmall{\textbf{-516.3(0)}} & \ssmall{\textbf{-517.8(0)}} \\
        \centering{\ssmall{HF 1k}} & \ssmall{\textbf{-15.0(0)}} & \ssmall{\textbf{-40.0(0)}} & \ssmall{-64.9(0)} & \ssmall{-109.3(0)} & \ssmall{-455.3(0)} & \ssmall{-496.7(0)} \\
        \centering{\ssmall{HF 30k}} & \ssmall{\textbf{-15.0(0)}} & \ssmall{\textbf{-40.0(0)}} & \ssmall{\textbf{-64.9(0)}} & \ssmall{-111.6(0)} & \ssmall{-478.3(0)} & \ssmall{-483.9(0)} \\
        \centering{\ssmall{HF 1M}} & \ssmall{\textbf{-15.0(0)}} & \ssmall{\textbf{-40.0(0)}} & \ssmall{\textbf{-64.9(0)}} & \ssmall{-116.9(0)} & \ssmall{-458.2(0)} & \ssmall{-468.9(0)} \\
        \centering{\ssmall{HF 32M}} & \ssmall{\textbf{-15.0(0)}} & \ssmall{\textbf{-40.0(0)}} & \ssmall{\textbf{-64.9(0)}} & \ssmall{-104.0(0)} & \ssmall{-493.0(0)} & \ssmall{-482.4(0)} \\
    \end{tabular}
    \caption{Best solution score found by quantum optimization methods executed on QPU (top) and by classical optimization methods executed on CPU (bottom).}
    \label{tab:best_scores}
\end{table}

\begin{figure}
    \centering
    \includegraphics[]{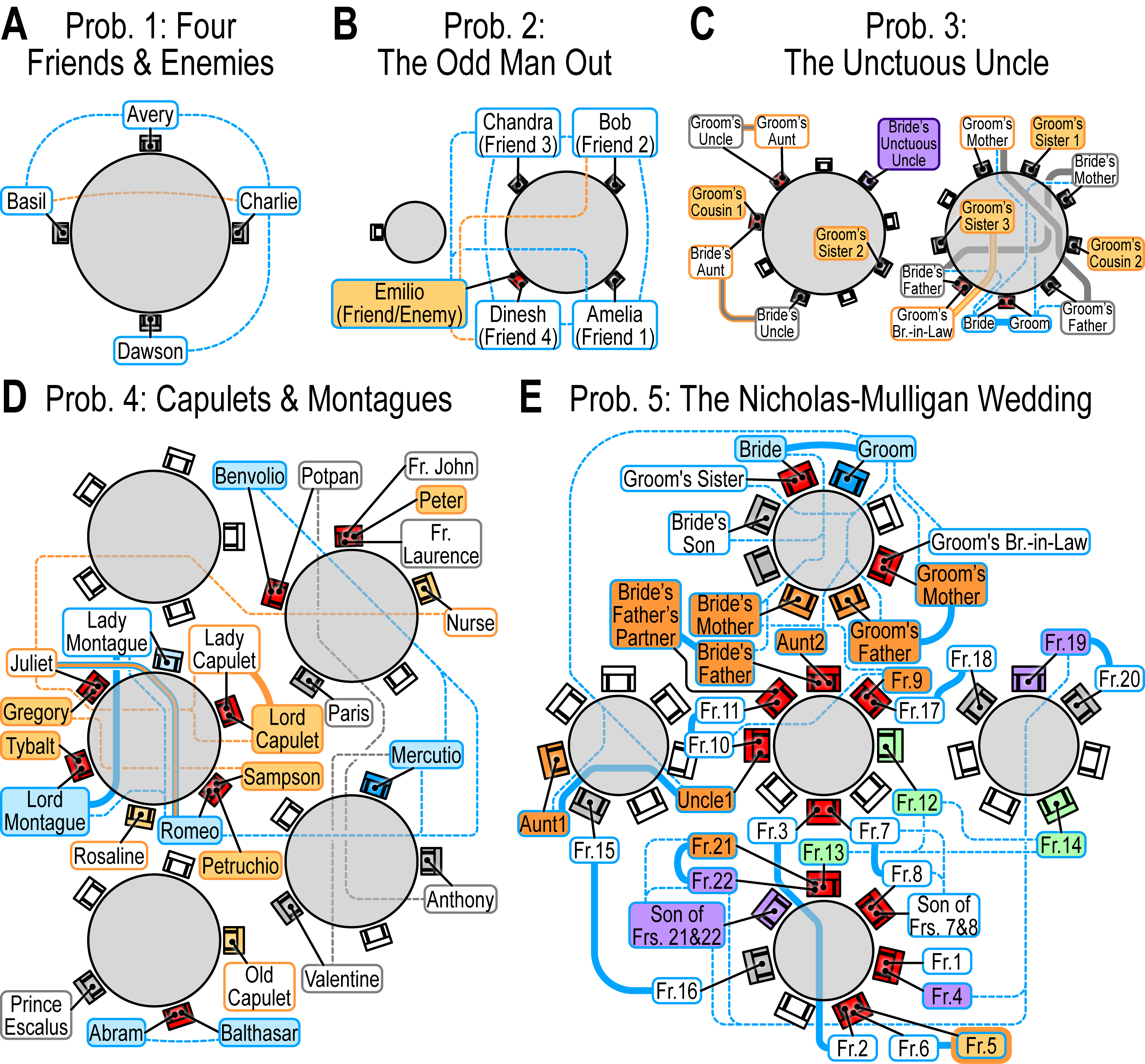}
    \caption{Seating solutions from quantum annealing.  (\textbf{A}) The quantum annealer robustly found all eight solutions to Prob. 1.  (\textbf{B}) In Prob. 2, the four friends were properly seated next to one another (blue dashed lines), but the ``odd man out'' (red) was seated atop one of his enemies.  (\textbf{C}) In Prob. 3, the ``unctuous uncle'' (purple) was isolated from those whom he might harass (orange), albeit at the cost of piling guests in seats far from him (red) and leaving the seats next to him empty. Couples were often not seated next to one another (thick lines).  (\textbf{D}) In Prob. 4, many guest overlaps were observed, including between Montagues (blue) and Capulets (orange).  Many proximity constraints were also violated (dashed lines), and one table was left empty.  (\textbf{E}) In Prob. 5s (our own wedding), many guest overlaps were also observed (red).  Several groups with shared interests (green, purple) were separated, though some couples (thick lines) were seated together, and offspring were often seated with their parents (dashed blue lines).  High-risk individuals (solid orange) were mostly kept far from a guest unable to receive a COVID-19 booster (pale orange).}
    \label{fig:quantum_results}
\end{figure}

    We found that one-hot and domain-wall encodings resulted in extraordinarily poor performance on even the simplest of these problems: one million samples produced no valid solutions (``NVS'' in \textbf{Table \ref{tab:best_scores}}) to the ``four friends and enemies'' (Prob. 1) problem. That is, there was no bitstring produced with proper one-hot or domain-wall encoding of an assignment of one seat per guest.  Rather than continuing to benchmark these mappings, we focused our effort on the approximate binary mapping.  This robustly found all eight solutions to the first problem (\textbf{Fig. \ref{fig:quantum_results}A}).  Indeed, we found that as few as 50 samples produced all eight solutions.  However, the slightly more complicated ``odd man out'' problem (Prob. 2), and all more complicated problems, resulted in no seat assignments that did not involve at least two guests being assigned to the same seat.  For the ``odd man out'' and ``unctuous uncle'' problems (Probs. 2 and 3), additional sampling allowed somewhat better-scoring solutions to be found, albeit not with fewer instances of multiple guests assigned to the same seat.  For the ``Capulets \& Montagues'' problem and the simplified version of our own wedding seating problem (Probs. 4 and 5s) more sampling did \textit{not} result in better solutions.  Representative failure modes for Probs. 2, 3, 4, and 5s are shown in \textbf{Fig. \ref{fig:quantum_results}B-E}.  In the case of the largest problem, our wedding's unsimplified seating optimization problem, \texttt{minor\_miner} was unable to find an embedding (``NE'' in \textbf{Table \ref{tab:best_scores}}).  Although intended for spatially localized optimization problems, tests with the variant \texttt{layout} embedding algorithm showed no improvement, nor any greater ability to find embeddings for larger problems.

\begin{figure}
    \centering
    \includegraphics[]{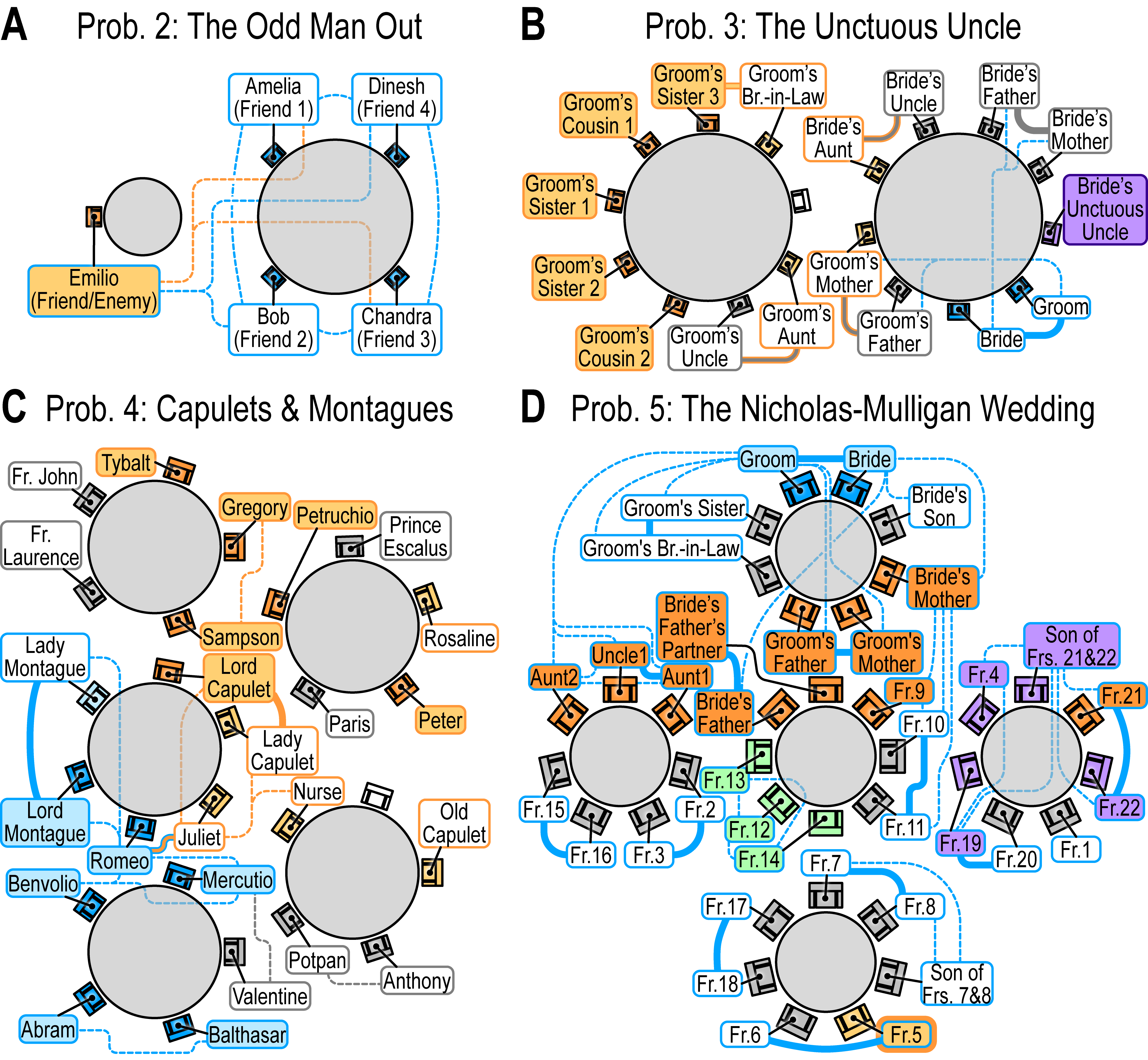}
    \caption{Seating solutions from classical Monte Carlo optimization.  (\textbf{A}) In Prob. 2, the ``odd man out'', who is enemies (orange dashed lines) with two of the four friends (blue dashed lines) and friends with the other two, is seated alone. (\textbf{B}) In Prob. 3, the bride's ``unctuous uncle'' (purple) is seated far from the groom's sisters and cousins (solid orange), and somewhat far from the groom's mother, aunts, and hotheaded brother-in-law (orange outlines), subject to the constraint that couples sit together (heavy lines) and the restraint that parents of the married couple are at their table (dashed blue lines). (\textbf{C}) In Prob. 4, the optimizer manages to seat Capulets (orange) and Montagues (blue) far apart, while keeping couples adjacent (heavy lines) and friends and family together (dashed lines), subject to head table restraints and constraints separating servants and nobles.  The most pugilistic individuals (solid/darker colours) are most separated.
    (\textbf{D}) For our own wedding reception, classical optimization successfully seated couples together (heavy lines), kept friends and family close to one another (dashed lines), grouped people with commonalities (\textit{e.g.} French speakers, green; scientists and enthusiasts, purple), and kept high-risk individuals (deep orange) far from a guest unable to receive a COVID-19 booster (pale orange), all subject to restraints placing the authors (blue seats) and close family at the head table.  (We used this classical solution at our wedding reception.)}
    \label{fig:classical_results}
\end{figure}

    In contrast, Masala's default Monte Carlo CFN optimizer performed well on these problems, consistently finding solutions in which each guest was assigned a unique seat, and satisfying most to all constraints.  For Prob. 1, even a 1,000-step trajectory found all globally optimal solutions, and for Prob. 2, a 1,000-step trajectory found 7 of 8 globally optimal solutions.  The solution found for Prob. 3 was also unchanged with more sampling, and was the best of any method, suggesting convergence to the global optimum even with only 1,000 steps.  Probs. 4, 5s, and 5 likely did not converge to the global optimum, since more sampling consistently found better and better results, but the solutions that were found were very good: in the ``Capulets \& Montagues'' problem, the two families were largely separated, and in our own wedding, the seating plans were all reasonable.  Representative examples produced with the Monte Carlo CFN optimizer are shown in \textbf{Fig. \ref{fig:classical_results}}.

    Although the experimental, classical hill-flattening Monte Carlo CFN optimizer showed better performance than the quantum annealing-based methods, reliably finding the best solution for Probs. 1 and 2, and the apparent optimum found by the default Masala Monte Carlo CFN optimizer for Prob. 3, its performance fell off for larger problems.  The hill-flattening optimizer was consistently able to find solutions in which guests did not overlap, but these solutions were noticeably worse than those produced by the default Monte Carlo optimizer given the same trajectory length.  Concerningly, these benchmarks revealed that the adaptive hill-flattening appeared to produce \textit{worse} results with longer trajectories as problems grew large. See Section \ref{subsec:discussion_what_benchmarks_reveal} for discussion of this phenomenon.

\section{Discussion}

\subsection{A Masala library for benchmarking classical and quantum optimizers}

    This work introduces the Seating Optimization Masala Plugins library, a C\texttt{++} library that converts seating problems into CFN problems solvable by any plugin Masala CFN optimizer.  This allows these problems to serve as a useful benchmark set for current and future CFN optimizers, implemented to run on classical or quantum hardware.
    Since seating problems yield CFN problems spanning sizes ranging from trivial to extremely difficult, they are particularly useful optimization benchmarks.
    CFN optimization is applicable to problems in many fields, including protein design and multibody docking problems in biomolecular modelling \cite{mulligan_designing_2019,pandey_multibody_2022}.
    Since optimizers need to be general enough to tackle problem instances with unforeseen properties, the development of software infrastructure to allow diverse problem types to be benchmarked with the same classical or quantum optimizer is crucial.  Because the Seating Optimization Masala Plugins library links Masala's plugin optimizers in the same way that protein design pipelines like Rosetta do \cite{zaborniak_open-source_2025-1}, we can now use the same optimizers to seat guests, to design proteins, or to dock molecules, eliminating discrepancies that could otherwise arise from separately implementing optimization approaches in benchmarking contexts and production pipelines.
    
    Beyond quantum annealing, there is active interest in developing gate-based quantum algorithms for protein design or multibody docking, using the quantum approximate optimization algorithm (QAOA) \cite{farhi_quantum_2014}, quantum imaginary-time evolution (QITE) \cite{motta_determining_2020}, or other ground state-seeking approaches.  As Masala plugins for these are created, our Seating Optimization Masala Plugins library will aid benchmarking.

\subsection{Seating optimization reveals limitations of current classical and quantum optimizers}
\label{subsec:discussion_what_benchmarks_reveal}
    
    The D-Wave 2000Q optimizer proved effective for peptide design using a one-hot CFN encoding \cite{mulligan_designing_2019}.  Recently, we have found that the Advantage 2 system is effective for protein design using our approximate binary CFN encoding (manuscript in preparation and \cite{zaborniak_toward_2025}).  We were therefore surprised to find that the Advantage 2 system struggles to solve any but the simplest seating optimization problems.  It had great difficulty satisfying the guest overlap constraint, frequently assigning multiple guests to the same seat.  Softer constraints, such as Gaussian proximity constraints, were better satisfied.  We suspect that solution quality is most influenced by embedding quality, and we hypothesize that large changes in performance from run to run were due to using different embeddings, masking any improvement from increasing the sampling in Probs. 4 and 5.  Unfortunately, short tests of the alternative \texttt{layout} embedding algorithm did not improve results.  Full understanding of this phenomenon will require repeated large-scale runs testing many embeddings, requiring investment of computing time beyond the scope of this short study.  Regardless, seating benchmarks will help to evaluate how new embeddings, mappings, and hardware improve performance.

    The development of quantum optimizers inevitably parallels the development of classical alternatives (sometimes quantum-inspired), so having comparable means of benchmarking both is crucial.  This benchmark set was useful for revealing problems with a classical hill-flattening Monte Carlo optimizer under development.  Loosely inspired by quantum tunnelling, this classical optimizer adaptively alters the loss function to reduce barrier heights based on the lowest-scoring states encountered, meaning that the past trajectory influences the search efficiency of the remainder.  
    While the intention is for this to \textit{improve} search efficiency, poorly tuned optimizer hyperparameters can cause the interplay of the annealing schedule and the accumulation of past state to \textit{worsen} search efficiency, which appears to be the case at present: for the largest problems, longer simulated annealing trajectories showed slightly worse performance.
    This problem can now be addressed, with new hyperparameter combinations checked against this benchmark.

    In contrast, the default Monte Carlo optimizer in the Standard Masala Plugins library performed well, finding high-quality seating optimization solutions that permitted real-world use at our own wedding.

\subsection{Seating optimization as a real-world application for classical and quantum CFN optimizers}

    Widespread adoption of quantum computers requires broadly understandable, real-world use cases.  As our own wedding demonstrated, seating optimization is a difficult yet commonplace task that benefits from advanced optimization approaches.  The Seating Optimization Masala Plugins library's input format permits new, real-world seating problems to be solved with any Masala classical or quantum CFN optimizer.  Just as we mitigated the spread of COVID-19 at our wedding, the inclusion of geometric constraints permits inclusion of guest health and safety considerations in the optimization problem.  Seating optimization is also isomorphic to related problems in healthcare settings, such as assignment of patients to hospital beds or to long-term care rooms, or the allocation of finite resources (equipment, staff, \textit{etc.}) to patients.  Rapid optimization strategies are particularly necessary for this last, to deal with changing circumstances.  Beyond providing an example problem class for testing optimization approaches, our methods can also assist in these situations.

\section{Conclusion}

    Our past success in mapping the peptide and protein design problem to quantum annealers, and in using the current D-Wave hardware for non-trivial macromolecular design problems, had led us to imagine that wedding seating optimization would be well within the capabilities of current-generation quantum computers.  It was therefore surprising and interesting to find that even seating optimization problems simple enough to solve by inspection, such as the ``odd man out'' problem (Prob. 2 in this work), seem to be beyond the abilities of current quantum optimizers using the same mappings that worked well for peptide and protein design.  This meant that what had started as a whimsical exercise in trying to seat the guests for our wedding using a quantum annealer resulted in the inadvertent construction of a useful set of test cases illustrating a class of problems that current classical optimizers solve robustly (so much so that we were able to seat our guests with Masala's classical Monte Carlo CFN optimizer), yet which quantum optimizers can barely tackle at all.  It is our hope that seating optimization as a class of problems, and the particular example problems that we devised, will aid the future development of classical and quantum optimization algorithms and hardware.  We make these problems, and the code to convert them to CFN problems and to run them with any plugin Masala optimizer (harnessing CPU or QPU), available under a permissive open-source licence to allow easy development and testing.  It is also our hope to allow real-world seating optimization problems, and other similarly-structured problems relevant to optimal personnel or resource allocation, to be solved using the best optimization methods available, be they classical or quantum approaches.  In our own case, our optimized seating plan ensured that a good time was had by all at our wedding reception.

\section*{Acknowledgements}

    \scriptsize{
    K.A.N. is funded by the Foundation for Health Care Quality.  V.K.M. is funded by the Simons Foundation.  The authors thank Tristan Zaborniak for helpful conversations, the Flatiron Institute’s Scientific Computing Core for ongoing support, and our friends and families for putting our optimization solution to the test at our wedding.  We also thank Nicole Grogan, our wedding planner, for accommodating our unconventional approach to seating. The computations reported here were performed using resources made available by the Flatiron Institute, and with D-Wave quantum annealer time funded by the Simons Foundation. The Flatiron Institute is a division of the Simons Foundation.
    }

%\section*{References}
\bibliographystyle{unsrt}
\bibliography{2_WeddingSeating_bibliography}

@article{zaborniak_open-source_2025-1,
        title = {The open-source {Masala} software suite: {Facilitating} rapid methods development for synthetic heteropolymer design},
        volume = {723},
        issn = {1557-7988},
        shorttitle = {The open-source {Masala} software suite},
        doi = {10.1016/bs.mie.2025.09.015},
        language = {eng}, 
        journal = {Methods in Enzymology},
        author = {Zaborniak, Tristan and Azadvari, Noora and Zhu, Qiyao and Turzo, S. M. Bargeen A. and Hosseinzadeh, Parisa and Renfrew, P. Douglas and Mulligan, Vikram Khipple},
        year = {2025},
        pages = {299--426},
}

@article{chancellor_domain_2019,
        title = {Domain wall encoding of discrete variables for quantum annealing and {QAOA}},
        volume = {4},
        issn = {2058-9565},
        url = {https://dx.doi.org/10.1088/2058-9565/ab33c2},
        doi = {10.1088/2058-9565/ab33c2},
        language = {en}, 
        number = {4},
        urldate = {2025-06-11},
        journal = {Quantum Science and Technology},
        publisher = {IOP Publishing},
        author = {Chancellor, Nicholas},
        month = aug,
        year = {2019},
        pages = {045004},
}

@article{leach_exploring_1998,
        title = {Exploring the conformational space of protein side chains using dead-end elimination and the {A}* algorithm},
        volume = {33},
        issn = {0887-3585},
        doi = {10.1002/(sici)1097-0134(19981101)33:2<227::aid-prot7>3.0.co;2-f},
        language = {eng},
        number = {2},
        journal = {Proteins},
        author = {Leach, A. R. and Lemon, A. P.},
        month = nov,
        year = {1998},
        pages = {227--239},
}

@article{kuhlman_native_2000,
        title = {Native protein sequences are close to optimal for their structures},
        volume = {97}, 
        issn = {0027-8424},
        url = {https://www.ncbi.nlm.nih.gov/pmc/articles/PMC27033/},
        number = {19},
        urldate = {2019-08-28},
        journal = {Proceedings of the National Academy of Sciences of the United States of America},
        author = {Kuhlman, Brian and Baker, David},
        month = sep,
        year = {2000},
        pages = {10383--10388},
}

@incollection{ghodge_computational_2022,
        address = {Washington, DC},
        series = {{ACS} {Symposium} {Series}},
        title = {Computational {Design} of {Peptide}-{Based} {Binders} to {Therapeutic} {Targets}},
        volume = {1417},
        isbn = {978-0-8412-9761-6 978-0-8412-9760-9},
        url = {https://pubs.acs.org/doi/abs/10.1021/bk-2022-1417.ch003},
        doi = {10.1021/bk-2022-1417.ch003},
        language = {en},
        urldate = {2022-09-01},
        booktitle = {Approaching the {Next} {Inflection} in {Peptide} {Therapeutics}: {Attaining} {Cell} {Permeability} and {Oral} {Bioavailability}},
        publisher = {American Chemical Society},
        author = {Mulligan, Vikram Khipple and Hosseinzadeh, Parisa},
        editor = {Ghodge, Swapnil V. and Biswas, Kaustav and Golosov, Andrei A.},
        month = aug,
        year = {2022},
        pages = {55--102},
}

@article{farhi_quantum_2014,
        title = {A {Quantum} {Approximate} {Optimization} {Algorithm}},
        url = {http://arxiv.org/abs/1411.4028},
        urldate = {2021-08-04},
        journal = {arXiv:1411.4028 [quant-ph]},
        author = {Farhi, Edward and Goldstone, Jeffrey and Gutmann, Sam},
        month = nov,
        year = {2014},
        note = {arXiv: 1411.4028},
}

@article{motta_determining_2020,
        title = {Determining eigenstates and thermal states on a quantum computer using quantum imaginary time evolution},
        volume = {16},
        copyright = {2019 The Author(s), under exclusive licence to Springer Nature Limited},
        issn = {1745-2481},
        url = {https://www.nature.com/articles/s41567-019-0704-4},
        doi = {10.1038/s41567-019-0704-4},
        language = {en},
        number = {2},
        urldate = {2026-04-09},
        journal = {Nature Physics},
        publisher = {Nature Publishing Group},
        author = {Motta, Mario and Sun, Chong and Tan, Adrian T. K. and O’Rourke, Matthew J. and Ye, Erika and Minnich, Austin J. and Brandão, Fernando G. S. L. and Chan, Garnet Kin-Lic},
        month = feb,
        year = {2020},
        pages = {205--210},
}

@misc{pandey_multibody_2022,
        title = {Multibody molecular docking on a quantum annealer},
        url = {http://arxiv.org/abs/2210.11401},
        doi = {10.48550/arXiv.2210.11401}, 
        urldate = {2022-10-31}, 
        publisher = {arXiv},
        author = {Pandey, Mohit and Zaborniak, Tristan and Melo, Hans and Galda, Alexey and Mulligan, Vikram K.},
        month = oct,
        year = {2022},
        note = {arXiv:2210.11401 [q-bio]},  
}

@misc{cai_practical_2014,
        title = {A practical heuristic for finding graph minors},
        url = {http://arxiv.org/abs/1406.2741},
        doi = {10.48550/arXiv.1406.2741},
        urldate = {2026-04-03},
        publisher = {arXiv},
        author = {Cai, Jun and Macready, William G. and Roy, Aidan},
        month = jun,
        year = {2014},
        note = {arXiv:1406.2741 [quant-ph]},
}

@article{adame_inhomogeneous_2020,
        title = {Inhomogeneous driving in quantum annealers can result in orders-of-magnitude improvements in performance},
        volume = {5},
        issn = {2058-9565},
        url = {https://doi.org/10.1088/2058-9565/ab935a},
        doi = {10.1088/2058-9565/ab935a},
        language = {en},
        number = {3},
        urldate = {2026-04-03}, 
        journal = {Quantum Science and Technology},
        publisher = {IOP Publishing},
        author = {Adame, Juan I and McMahon, Peter L},
        month = jun,
        year = {2020},
        pages = {035011},
}

@article{leman_macromolecular_2020,
        title = {Macromolecular modeling and design in {Rosetta}: recent methods and frameworks},
        volume = {17},
        copyright = {2020 Springer Nature America, Inc.},
        issn = {1548-7105},
        shorttitle = {Macromolecular modeling and design in {Rosetta}},
        url = {https://www.nature.com/articles/s41592-020-0848-2},
        doi = {10.1038/s41592-020-0848-2},
        language = {en},
        number = {7},
        urldate = {2025-01-07},
        journal = {Nature Methods},
        publisher = {Nature Publishing Group},
        author = {Leman, Julia Koehler and Weitzner, Brian D. and Lewis, Steven M. and Adolf-Bryfogle, Jared and Alam, Nawsad and Alford, Rebecca F. and Aprahamian, Melanie and Baker, David and Barlow, Kyle A. and Barth, Patrick and Basanta, Benjamin and Bender, Brian J. and Blacklock, Kristin and Bonet, Jaume and Boyken, Scott E. and Bradley, Phil and Bystroff, Chris and Conway, Patrick and Cooper, Seth and Correia, Bruno E. and Coventry, Brian and Das, Rhiju and De Jong, René M. and DiMaio, Frank and Dsilva, Lorna and Dunbrack, Roland and Ford, Alexander S. and Frenz, Brandon and Fu, Darwin Y. and Geniesse, Caleb and Goldschmidt, Lukasz and Gowthaman, Ragul and Gray, Jeffrey J. and Gront, Dominik and Guffy, Sharon and Horowitz, Scott and Huang, Po-Ssu and Huber, Thomas and Jacobs, Tim M. and Jeliazkov, Jeliazko R. and Johnson, David K. and Kappel, Kalli and Karanicolas, John and Khakzad, Hamed and Khar, Karen R. and Khare, Sagar D. and Khatib, Firas and Khramushin, Alisa and King, Indigo C. and Kleffner, Robert and Koepnick, Brian and Kortemme, Tanja and Kuenze, Georg and Kuhlman, Brian and Kuroda, Daisuke and Labonte, Jason W. and Lai, Jason K. and Lapidoth, Gideon and Leaver-Fay, Andrew and Lindert, Steffen and Linsky, Thomas and London, Nir and Lubin, Joseph H. and Lyskov, Sergey and Maguire, Jack and Malmström, Lars and Marcos, Enrique and Marcu, Orly and Marze, Nicholas A. and Meiler, Jens and Moretti, Rocco and Mulligan, Vikram Khipple and Nerli, Santrupti and Norn, Christoffer and Ó’Conchúir, Shane and Ollikainen, Noah and Ovchinnikov, Sergey and Pacella, Michael S. and Pan, Xingjie and Park, Hahnbeom and Pavlovicz, Ryan E. and Pethe, Manasi and Pierce, Brian G. and Pilla, Kala Bharath and Raveh, Barak and Renfrew, P. Douglas and Burman, Shourya S. Roy and Rubenstein, Aliza and Sauer, Marion F. and Scheck, Andreas and Schief, William and Schueler-Furman, Ora and Sedan, Yuval and Sevy, Alexander M. and Sgourakis, Nikolaos G. and Shi, Lei and Siegel, Justin B. and Silva, Daniel-Adriano and Smith, Shannon and Song, Yifan and Stein, Amelie and Szegedy, Maria and Teets, Frank D. and Thyme, Summer B. and Wang, Ray Yu-Ruei and Watkins, Andrew and Zimmerman, Lior and Bonneau, Richard},
        month = jul,
        year = {2020},
        pages = {665--680},
}

@incollection{leaver-fay_rosetta3_2011,
        title = {Rosetta3},
        volume = {487},
        isbn = {978-0-12-381270-4},
        doi = {10.1016/B978-0-12-381270-4.00019-6},
        language = {en},
        booktitle = {Methods in {Enzymology}},
        publisher = {Elsevier},
        author = {Leaver-Fay, Andrew and Tyka, Michael and Lewis, Steven M. and Lange, Oliver F. and Thompson, James and Jacak, Ron and Kaufman, Kristian W. and Renfrew, P. Douglas and Smith, Colin A. and Sheffler, Will and Davis, Ian W. and Cooper, Seth and Treuille, Adrien and Mandell, Daniel J. and Richter, Florian and Ban, Yih-En Andrew and Fleishman, Sarel J. and Corn, Jacob E. and Kim, David E. and Lyskov, Sergey and Berrondo, Monica and Mentzer, Stuart and Popović, Zoran and Havranek, James J. and Karanicolas, John and Das, Rhiju and Meiler, Jens and Kortemme, Tanja and Gray, Jeffrey J. and Kuhlman, Brian and Baker, David and Bradley, Philip},
        year = {2011},
        pages = {545--574},
}

@phdthesis{zaborniak_toward_2025,
        address = {Victoria, British Columbia, Canada},
        type = {Ph.{D}. dissertation},
        title = {Toward quantum computational biomolecular structure prediction},
        url = {https://hdl.handle.net/1828/22736},
        language = {English},
        urldate = {2026-01-18},
        school = {University of Victoria},
        author = {Zaborniak, Tristan},
        year = {2025},
}

@article{traore_new_2013,
        title = {A new framework for computational protein design through cost function network optimization},
        volume = {29},
        issn = {1367-4811},
        doi = {10.1093/bioinformatics/btt374},
        language = {eng},
        number = {17},
        journal = {Bioinformatics (Oxford, England)},
        author = {Traoré, Seydou and Allouche, David and André, Isabelle and de Givry, Simon and Katsirelos, George and Schiex, Thomas and Barbe, Sophie},
        month = sep,
        year = {2013},
        pages = {2129--2136},
}

@article{pierce_protein_2002,
        title = {Protein design is {NP}-hard}, 
        volume = {15},
        issn = {0269-2139},
        doi = {10.1093/protein/15.10.779},
        language = {eng}, 
        number = {10},
        journal = {Protein Engineering},
        author = {Pierce, Niles A. and Winfree, Erik},
        month = oct,
        year = {2002},
        pages = {779--782},
}

@misc{mulligan_designing_2019,
        title = {Designing {Peptides} on a {Quantum} {Computer}},
        copyright = {© 2019, Posted by Cold Spring Harbor Laboratory. The copyright holder for this pre-print is the author. All rights reserved. The material may not be redistributed, re-used or adapted without the author's permission.},
        url = {https://www.biorxiv.org/content/10.1101/752485v1},
        doi = {10.1101/752485},
        language = {en},
        urldate = {2019-12-11}, 
        journal = {bioRxiv},
		note={\textit{bioRxiv}, \href{https://doi.org/10.1101/752485}{doi:10.1101/752485}},
        author = {Mulligan, Vikram Khipple and Melo, Hans and Merritt, Haley Irene and Slocum, Stewart and Weitzner, Brian D. and Watkins, Andrew M. and Renfrew, P. Douglas and Pelissier, Craig and Arora, Paramjit S. and Bonneau, Richard},
        month = sep,
        year = {2019},
        pages = {752485},
}

\section*{Conflict of interest statement}

    \scriptsize{
    V.K.M. is a co-founder and shareholder of Menten AI, a peptide design company that uses advanced optimization algorithms.
    }

\end{document}